\documentclass[oneside,english]{amsart}
\usepackage[T1]{fontenc}
\usepackage[utf8]{luainputenc}
\usepackage{amsthm}
\usepackage{stackrel}

\makeatletter
\numberwithin{equation}{section}
\numberwithin{figure}{section}
\theoremstyle{plain}
\newtheorem{thm}{\protect\theoremname}
  \theoremstyle{definition}
  \newtheorem{defn}[thm]{\protect\definitionname}

\makeatother

\usepackage{babel}
  \providecommand{\definitionname}{Definition}
\providecommand{\theoremname}{Theorem}

\begin{document}

\title{Finite and infinite basis in P and NP}

\author{Koji KOBAYASHI}

\maketitle

\section{Abstract}

This article provide new approach to solve P vs NP problem by using
cardinality of bases function. About NP-Complete problems, we can
divide to infinite disjunction of P-Complete problems. These P-Complete
problems are independent of each other in disjunction. That is, NP-Complete
problem is in infinite dimension function space that bases are P-Complete.
The other hand, any P-Complete problem have at most a finite number
of P-Complete basis. The reason is that each P problems have at most
finite number of Least fixed point operator. Therefore, we cannot
describe NP-Complete problems in P. We can also prove this result
from incompleteness of P.

\section{Difference of basis between P and NP}

By using SAT and these verification, we prove that some NP-Complete
problems have infinite basis of P-Complete problems.
\begin{defn}
\label{def: Problems} We will use the term ``$v_{i}\in V$'' as
problem which verify formula with special valuation $i$. 

That is, if 

$t\in SAT$

then

$v_{i}\left(t\right)=\top\leftrightarrow t\left(i\right)=\top$\end{defn}
\begin{thm}
\label{thm: V is P-Complete} $v_{i}\in P-Complete$\end{thm}
\begin{proof}
First, we show that $v_{i}\in P$. A Polynomial DTM can verify valuation
$i$ to a given formula $f$ and accept if $f\left(i\right)=\top$.

Next, we show that $CIRCUIT-VALUE\leq_{L}v_{i}$. $CIRCUIT-VALUE\in P-Complete$\cite{Book1},
therefore if $CIRCUIT-VALUE\leq_{L}v_{i}\in P$ then $v_{i}\in P-Complete$.
If we modify $C\rightarrow C^{\prime}$ to match $x\rightarrow i$,
$v_{i}$ compute $C^{\prime}$ as $\left\langle C,x\right\rangle $.
We can modify $C\rightarrow C^{\prime}$ to negate some $C$ variables
that $x$ mismatch $i$. This modification can compute in $L$.

Therefore, $CIRCUIT-VALUE\leq_{L}v_{i}\in P$ and $v_{i}\in P-Complete$.\end{proof}
\begin{thm}
\label{thm: Basis of SAT} $V$ is basis of $SAT$ \end{thm}
\begin{proof}
To think about relation between $SAT$ and $v_{i}\in V$, $SAT$ is
disjunction of $V$, i.e.

$SAT=\bigcup V=\stackrel[i=0]{\infty}{\bigvee}v_{i}$

Each $v_{i}$ is independent of each other in disjunction because
every input $p$ have another input $q$ that change only $v_{i}$
output.

$\forall p\exists q\left(\left(v_{0}\left(p\right),\cdots,v_{i}\left(p\right),\cdots\right)\rightarrow\left(v_{0}\left(q\right)=v_{0}\left(p\right),\cdots,v_{i}\left(q\right)=\neg v_{i}\left(p\right),\cdots\right)\right)$

If $v_{i}\left(p\right)=\top$ then $q=p\wedge\left(\neg i\right)$

else if $v_{i}\left(p\right)=\bot$ then $q=p\vee\left(i\right)$

That is, $V\setminus\left\{ v_{i}\right\} $ cannot compute $SAT$
problems.

Therefore $V$ is basis of $SAT$.
\end{proof}
From descriptive complexity, $P=FO+LFP$\cite{Book1,Book2,Book3}.
This means that every P problem have at most a finite number of LFP
operators in finite first-order logic model. Therefore P problem have
at most a finite number of P-Complete basis.
\begin{thm}
\label{thm: Basis of P problem}Any $p\in P$ have at most a finite
number of P-Complete basis.\end{thm}
\begin{proof}
To prove it by using reduction to absurdity. We assume that $p\in P$
have infinite number of basis of P-Complete. These basis independent
of each other and have independent LFP operators. But $P=FO+LFP$
have at most finite number of LFP operators. Therefore we cannot describe
$p$ in finite length $FO+LFP$. \end{proof}
\begin{thm}
\label{thm: P is not NP}$P\neq NP$\end{thm}
\begin{proof}
Mentioned above \ref{thm: Basis of SAT}, $SAT$ have infinite P-Complete
basis. But mentioned above \ref{thm: Basis of P problem}, any $p\in P$
have finite P-Complete basis. Therefore $SAT$ is not any $p\in P$.
\end{proof}

\section{From view of countable and continuum}

We show another proof from the view of completeness.
\begin{thm}
\label{thm: P is not NP too}$P\neq NP$\end{thm}
\begin{proof}
Let $\left\langle v,i\right\rangle $ be a code number of $v_{i}$.
To assign this number after the decimal point, $0.\left\langle v,i\right\rangle $
correspond to number within $\left[0,1\right]$, and $\left[0,0.\left\langle v,i\right\rangle \right]+\bigcup V=\left[0,0.\left\langle v,i\right\rangle \right]+\stackrel[i=0]{\infty}{\bigvee}v_{i}$
correspond to Dedekind cut of $P$.

If $\left[0,0.\left\langle v,i\right\rangle \right]+\stackrel[i=0]{\infty}{\bigvee}v_{i}$
also $P$ then $P$ become isomorphic as real number and contradict
that $P$ is countable. Therefore $NP\ni\bigcup V\notin P$ and $P\neq NP$.\end{proof}

\end{document}